\documentclass[prb,showpacs,epsffig]{revtex4}

\usepackage{graphicx}
\usepackage{dcolumn}
\usepackage{bm}

\begin{document}

\title{Signature of small rings in the Raman spectra of normal and compressed amorphous silica:
A combined classical and {\it ab initio} study.}
\author{Abdelali Rahmani}
\affiliation{D\'epartement de Physique, Universit\'e MY Ismail,
Facult\'e des Sciences, BP 4010,50000 Mekn\`es, Morocco}
\author{Magali Benoit}
\affiliation{Laboratoire des Verres 
(UMR CNRS 5587), Universit\'e Montpellier II, 34095 Montpellier Cedex 5, France}
\author{Claude Benoit}
\affiliation{Groupe de Dynamique des
Phases Condens\'ees (UMR CNRS 5581), Universit\'e Montpellier II, 34095 Montpellier
Cedex 5, France}

\date{\today}

\begin{abstract}

We calculate the parallel (VV) and perpendicular (VH) polarized Raman spectra of amorphous
silica. Model SiO$_2$ glasses, uncompressed and compressed,
were generated by a combination of classical and \emph{ab initio} molecular-dynamics 
simulations and their dynamical matrices were computed
within the framework of the density functional theory.
 The Raman scattering intensities were determined using the bond-polarizability model
and a good agreement with experimental spectra was found. 
We confirm that the modes associated to the fourfold and 
threefold rings produce most of the Raman intensity of the D$_1$ and D$_2$ peaks, respectively,
in the VV Raman spectra.
Modifications of the Raman spectra upon compression
are found to be in agreement with experimental data. We show that the modes
associated to the fourfold rings still exist upon compression 
but do not produce a strong Raman intensity, whereas
the ones associated to the threefold rings do. This result strongly suggests that
the area under the D$_1$ and D$_2$ peaks is not directly proportional to 
the concentration of small rings in amorphous SiO$_2$.
\end{abstract}\vskip 1cm


\maketitle

\section{Introduction\newline}

Raman spectroscopy has provided a powerful tool to study the dynamical
properties of amorphous materials. The interpretation of the vibrational spectra of
vitreous silica (a-SiO$_2$) has been of interest for many years and considerable progress has been
made. Despite this fact, the origin of the D$_1$ and D$_2$ lines at 495 cm$^{-1}$
and 606 cm$^{-1}$, respectively, in the Raman spectrum is still not fully understood.
In order to explain the frequencies of these bands, Galeener {\it et al.} 
\cite{Galeener1,Galeener2,Galeener3,Barrio,Martin} has argued that D$_2$ can be assigned
to the breathing modes of a planar threefold ring, and D$_1$  to that of a
regular but slightly puckered fourfold ring within the continuous-random-network
(CRN)\cite{Zachariasen,Bell1}. On the other hand and within the submicrocrystallite
approach, Phillips \cite{Phillips1,Phillips2,Phillips3,Phillips4} has proposed that within
the a-SiO$_2$ structure there exist two types of non-bridging oxygen (NBO) sites at the
surface of the cluster, which can be associated to the structure responsible for the
 D$_1$ and D$_2$ defect lines. To date, much work has been done in the light of both models.
Using a pair potential and the bond-polarizability (BP) model \cite{Bell1,Bell2},
Murray and Ching \cite{Murray} have calculated the Raman spectra for normal and
compressed SiO$_2$ glass models based on  the CRN model with periodic boundary conditions. Their
results generally reproduce the shapes of the bands present in experimental Raman spectra,
particularly the D$_1$ and D$_2$ lines in contrast with the paracrystalline theory of
SiO$_2$. Moreover, the ring statistic analysis showed that there are no three or fourfold
rings in their models. This means that the D$_1$ and D$_2$ lines are not strictly due to the
threefold rings and fourfold rings as suggested by Galeener. To conciliate this later
point of view and their work, Murray and Ching indicated that the Si-O-Si angles close to
those found in the three and fourfold rings are responsible for these modes. More recently,
using molecular-dynamics (MD) simulations and the BP model, Zotov \emph{et al.}
\cite{Zotov} have calculated the Raman spectra of a-SiO$_2$ models. The obtained VV
Raman spectra were, generally, in good agreement with experiment but, despite the
presence of three and fourfold rings, their models failed to reproduce the D$_1$ and D$_2$
peaks. Using a first-principles density functional approach, Umari \emph{et al}
\cite{Umari} have calculated the Raman intensities of $\alpha$-quartz by evaluating the
variation of the polarizability tensors for finite displacements of the atoms. This work
supports the use of the bond polarizability model for the calculation of the Raman
intensities in a-SiO$_2$. By applying this method to an amorphous SiO$_2$ sample
containing a large concentration of threefold rings, they were able to relate the intensity
of the D$_1$ and D$_2$ peaks to the concentration of four and threefold rings, respectively \cite{Umari2}.
In a recent study, Benoit and Kob compared the vibrational properties of model SiO$_2$ glasses
generated by molecular-dynamics simulations using the effective force field by van Beest
\emph{et al} (BKS)\cite{vanBeest} with the ones obtained if the BKS structure is relaxed
using \emph{ab initio} calculations \cite{Benoit1}. It was found that this
relaxation significantly improves the agreement of the density of states with the
experimental result. Therefore it has become of interest to use the data of this
approach to determine the Raman spectra of these systems.\\

In this work, we present the polarized Raman spectra of compressed and uncompressed
a-SiO$_2$ samples prepared by combined classical and \emph{ab initio} molecular dynamics
simulations. Raman spectra were calculated using the bond-polarizability  model
 and  diagonalisation of the dynamical matrix, 
as obtained from the evaluation of the second derivatives of the total energy with
respect to atomic displacements by taking finite differences of the atomic forces.
In the following section,
we describe the models used to compute the Raman spectra. Then in the third section
we present the calculated spectra and a detailed analysis of the Raman signatures of
the small rings. Finally the spectra are also calculated for compressed silica
and a discussion in terms of the Raman signature of the rings is also given before the
conclusion.

\section{COMPUTATIONAL METHOD}

The time averaged power flux of Raman scattered light in a given direction, with a
frequency between $\omega_f$ and $\omega_f+d\omega_f$ in a solid angle $d\Omega$, is
related to the differential scattering cross-section:
\begin{equation}
\frac{d^2 \sigma }{d \Omega d \omega_f}=\frac{1}{8\pi^2 c^2}\omega^3_f
\omega_i(B(\omega)+1)\hbar\sum_{\alpha\beta\gamma\lambda} v_\alpha v_\beta
H_{\alpha\gamma\beta\lambda}(\omega) w_\gamma w_\lambda,
\end{equation}
where $\omega= \omega_i - \omega_f$ and 
\begin{equation}
H_{\alpha\gamma\beta\lambda}(\omega)=\sum_{j}a^\star_{\alpha\gamma}(j)a_{\beta\lambda}(j
)\frac{1}{2 \omega_j}(\delta(\omega-\omega_j)-\delta(\omega+\omega_j)),
\end{equation}
with
\begin{equation}
a_{\alpha\gamma}(j)=\sum_{n \delta}\frac{\pi^n_{\alpha\gamma,\delta}}{\sqrt{M_n}}<n \delta
\mid j>,
\label{eq}
\end{equation}
$\omega_i$ is the frequency of the incident light, $\vec{v}$ and $\vec{w}$ are the
polarization unit vectors for scattered and incident light respectively, $B(\omega)$ is
the Bose factor, $M_n$ the mass of the $n$th atom, $\omega_{j}$ and $<n \delta \mid j>$
are respectively the frequency and the $(n \delta)$ component of the $j$th mode. The Greek characters
denote the Cartesian components (x,y,z).
The coefficients ${\pi^n_{\alpha\gamma,\delta}}$ connect the polarization fluctuations
to the atomic motions \cite{rahmani,viliani} and they are obtained by expanding the
polarizability tensor $\widetilde{\pi}^n$ in terms of atomic displacements $u^n_\delta$,
with
\begin{equation}
\pi^n_{\alpha\gamma,\delta}=\sum_{m} \left.\left(\frac{\partial\pi^m_{\alpha\gamma}}{\partial
u^n_\delta}\right) \right|_{\{ u^n_{\delta} \} = 0} \ .
\end{equation}
In addition to these coefficients, the calculation of Raman intensities requires mode
frequencies and eigenvectors: $\omega_{j}$ and $< n \delta  \mid j>$, obtained by solving
the equation:
\begin{equation}
\tilde{D} \mid j>=\omega^2_j \mid j >,
\end{equation}
where $\tilde{D}$ is the dynamical matrix of the system, with the element
$D_{\alpha\beta}(nm)=<n \alpha \mid\tilde{D}\mid m \beta >$.
Each element of the dynamical matrix, $\tilde{D}$, is given by:
\begin{equation}
D_{\alpha\beta} (n,m)=\frac{1}{\sqrt{M_n M_m}}\phi_{\alpha\beta}(n,m),
\end{equation}
with $\phi_{\alpha\beta}(n,m)$ being the force constants between atoms $n$ and $m$. In
our calculations, the interactions are given by a first-principles approach.\\

If we assume that scattering can be described within the framework of the bond polarizability (BP)
theory, the polarization is only modulated by nearest-neighbor bonds and the components
of the induced polarizability tensor $\tilde{\pi}$ are given by \cite{Bell2}
 \begin{equation}
\pi_{\alpha\beta}(r)=\frac{1}{3}(\alpha_l+2
\alpha_p)\delta_{\alpha\beta}+(\alpha_l-\alpha_p)(\hat{r}_\alpha
\hat{r}_\beta-\frac{1}{3} \delta_{\alpha\beta}),
\end{equation}
where  $\hat{r}$ is the unit
vector along vector $\vec{r}$ which connects the $n$ and $m$ atoms linked by the bond.
The parameters $\alpha_{l}$ and $\alpha_{p}$ correspond to the longitudinal and
perpendicular bond polarizability, respectively. Within this approach and in order to
evaluate the derivatives, one assumes that the bond polarizability parameters are
functions of the bond lengths $r$ only. The derivatives $\pi^n_{\alpha\beta,\gamma}$ are then 
given by \cite{Bell2}
\begin{equation}
\pi^n_{\alpha\beta,\gamma}=\sum_{m}\frac{1}{3}(2\alpha'_p+\alpha'_l)\delta_{\alpha\beta}
\hat{r}_\gamma+(\alpha'_l-\alpha'_p)(\hat{r}_{\alpha}\hat{r}_{\beta}-\frac{1}{3}
\delta_{\alpha\beta})\hat{r}_{\gamma}+\frac{(\alpha_l-\alpha_p)}{r}(\delta_{\alpha\gamma}
\hat{r}_\beta+\delta_{\beta\gamma}\hat{r}_\alpha -2 \hat{r}_\alpha \hat{r}_\beta \hat{r}_\gamma)
\end{equation}
where $\alpha'= \left.(\partial\alpha/\partial r) \right|_{r=r_0}$, $r_0$ is the equilibrium distance. \\
In our a-SiO$_2$ model, a single type of bond occurs (Si-O) and the bond polarizability
model is completely defined by three parameters: $\bar{\alpha}=(2\alpha'_{p}+\alpha'_{l})$,
$\bar{\beta}=\alpha'_{l}-\alpha'_{p}$  and $\bar{\gamma}=(\alpha_{l}-\alpha_{p})/r$.   \\

\section{Raman scattering in amorphous SiO$_2$}
\label{sec:sio2}

The amorphous SiO$_2$ samples used in this study contain 26 SiO$_2$ units in a cubic box
with periodic boundary conditions at the experimental density of 2.2 g.cm$^{-3}$.
They were generated by molecular-dynamics simulations using the van Beest {\it et al.}
force field \cite{vanBeest}: Three well-equilibrated liquids at 3000 K were quenched
to 300 K using three different quench rates
($7 \cdot 10^{10}$ K/s (A), $3 \cdot 10^{11}$ K/s (B) and $5 \cdot 10^{12}$ K/s (C)).
The structures of the model samples were then refined using first-principles calculations of
the Car-Parrinello type at 300 K using the CPMD code \cite{CPMD} 
(for the simulation details, see Ref. \cite{Benoit1}).
After a relaxation of the atomic positions to 0 K in the density functional theory framework,
the dynamical matrices of these samples were computed
by evaluating the second derivatives of the {\it ab initio} total energy with respect to
the atomic displacements by taking the finite differences of the atomic forces.
The vibrational frequencies and corresponding modes were obtained by 
diagonalisation of the dynamical matrices. 

The structural characteristics of the three samples were first analyzed in terms of the
pair correlation functions and of the static structure factors. All samples present a perfect
tetrahedrally connected network and no noticeable sample specificities were found
on such averaged quantities. \\

\subsection{Density of states}

\begin{figure}\vskip 7mm
\centerline{\includegraphics{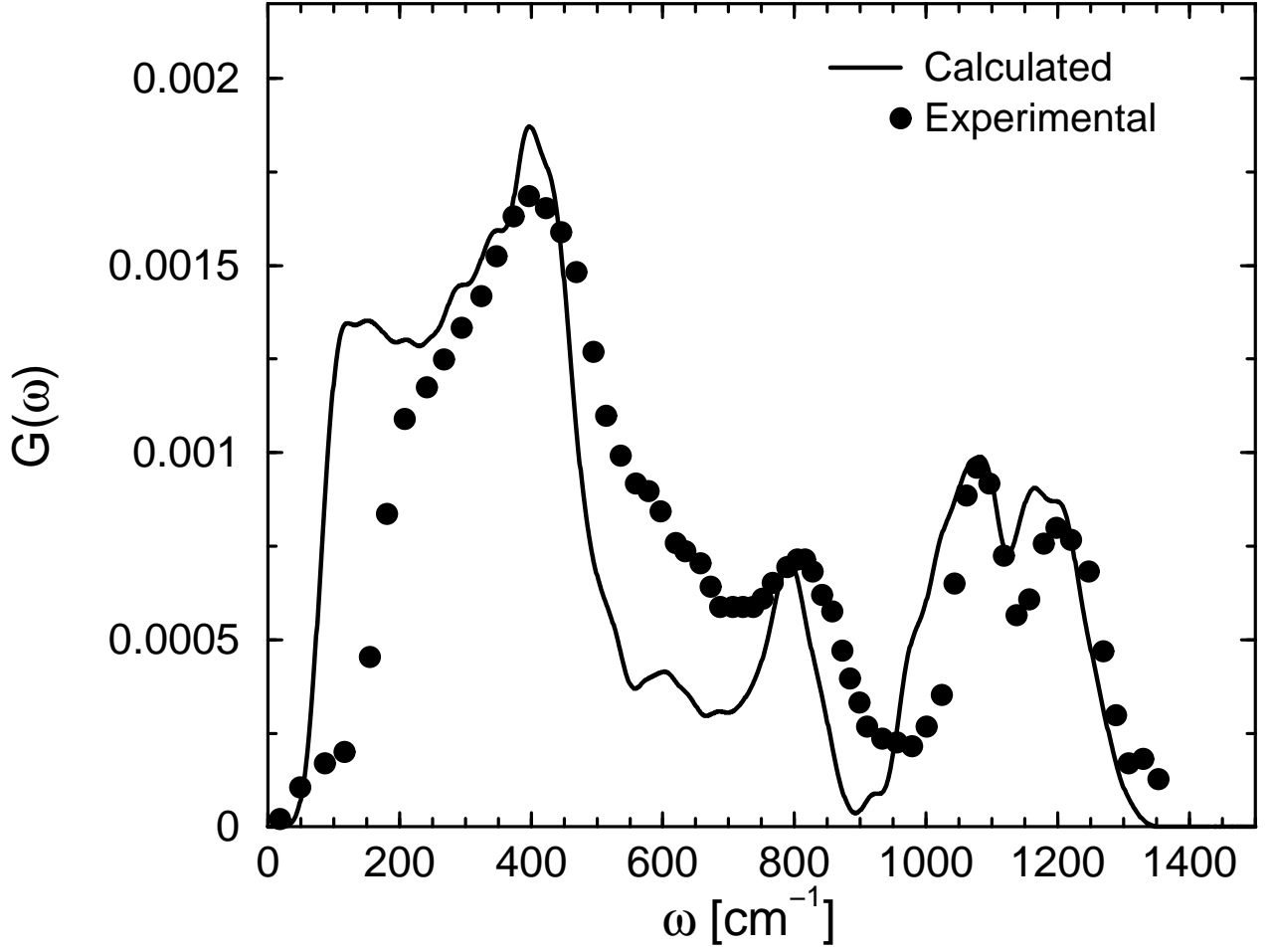}}
\caption{Frequency dependence of the effective density of states $G(\omega)$ obtained
from {\it ab initio} (solid line) simulations and compared to neutron scattering 
experiments from Ref. \protect\cite{Carpenter} (filled circles). The calculated $G(\omega)$
has been obtained using the experimental values for the neutron scattering length factors 
$b_{Si}$=4.149 fm and $b_O$=5.803 fm and a Gaussian broadening of width $2\sigma$= 35 cm$^{-1}$.}
\label{fig:vdos}
\end{figure}

The vibrational properties of a-SiO$_2$ have been extensively investigated by many
experimental and some theoretical means. In order to show the validity of the
vibrational features of our model, we report, in Figure \protect\ref{fig:vdos},
the calculated  effective neutron scattering cross
section $G(\omega)=C(\omega)g(\omega)$, where $g(\omega)$ is the vibrational density of states
and $C(\omega)$ was computed using the incoherent approximation, following Ref. \cite{Taraskin}.
The calculated $G(\omega)$ was obtained by averaging over the $G(\omega)$ of the three
SiO$_2$ samples A, B and C. The calculated $G(\omega)$ is compared to the experimental one \cite{Carpenter}
and a good agreement between the two is observed.
Shapes and positions of the principal peaks are well reproduced with a low-frequency band
at 420 cm$^{-1}$, an intermediate frequency band at  790 cm$^{-1}$ and a double peak at
1050 cm$^{-1}$  and 1195 cm$^{-1}$ in the high-frequency region \cite{Murray,Zotov,vanBeest}. 
Note that the lack of a small peak at around 4 THz in the experimental data is  related to the
insufficient experimental resolution ~\cite{wischnewski98}. \\

\subsection{Raman spectra}

The purpose of the present subsection is to present our calculation results for the polarized
parallel (VV) and perpendicular (VH) Raman spectra of different samples of a-SiO$_2$. 
We assumed that scattering was produced by the BP mechanism. 
In all spectra, intensities represent the so-called reduced intensities, 
which are obtained by multiplying the experimentally
measured spectrum with the correction factor $(\omega/(B(\omega)+1))$. In the VV
configuration, both the incident and scattered polarizations are along the z axis and,
for the VH configuration, the incident and scattered polarizations are along the z and x
axis respectively. The intensities of the different spectra are normalized and,
to make the comparison with the experimental results more realistic,
we averaged over all sample orientations in the Raman spectra calculations. \\

\begin{figure}
\centerline{\includegraphics{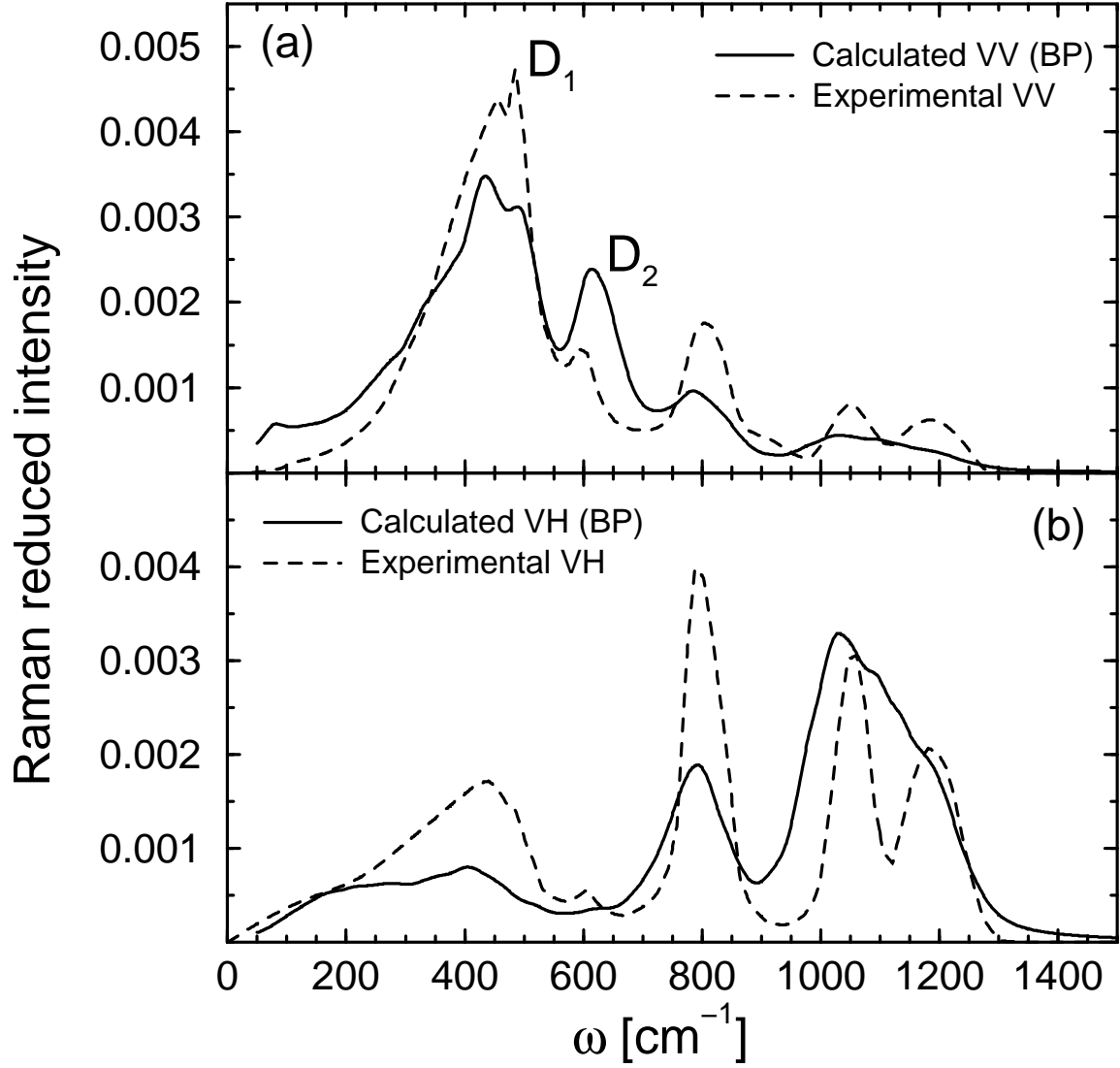}}
\caption{{\bf (a):} Frequency dependence of the VV 
Raman intensity calculated using the {\it ab initio}
eigen modes and the bond polarizability model (solid line). The dashed line is the 
experimental spectrum from Ref. \protect\cite{Galeener2}.
{\bf (b):}  Frequency dependence of the VH
Raman intensity calculated using the {\it ab initio}
eigen modes and the bond polarizability model (solid line). The dashed line is the 
experimental spectrum from Ref. \protect\cite{Galeener2}.
The calculated and experimental
spectra have been normalized to the same total area.
In the calculated spectra, the line shape is assumed to be Lorentzian and the line width is
fixed to 25 cm$^{-1}$.  }
\label{fig:vvhBP}
\end{figure}

Firstly the Raman spectra were calculated 
 for normal density (2.20 g/cm$^3$) a-SiO$_2$ samples. The empirical
parameters used in this study are those derived from the results of Umari \emph{et al.} \cite{Umari}
for $\alpha$-quartz: $\bar{\alpha}$=1.0, $\bar{\beta}$=0.25 and $\bar{\gamma}$=0.07. The
calculated Raman spectra (solid lines), which are obtained by an average over the
three different a-SiO$_2$ samples, are plotted together with the experimental
result of Galeener  \cite{Galeener2} in Figure \protect\ref{fig:vvhBP} (dashed lines).

The calculated VV Raman spectrum (Figure \protect\ref{fig:vvhBP}(a)) 
is dominated by a lower band, around 500 cm$^{-1}$
and reproduces the D$_1$ = 495 cm$^{-1}$ and D$_2$ = 606 cm$^{-1}$ peaks found in
experimental spectra \cite{Galeener2,Phillips4,McMillan,Stolen}. The calculated D$_1$ and D$_2$
occur at 489 cm$^{-1}$ and 614 cm$^{-1}$, respectively. For these bands, the curve is
very close in peak positions and qualitatively in peak shapes to the experimental curve.
 We note also that
the bands at 436 cm$^{-1}$  and 790 cm$^{-1}$ are close to the experimental bands
observed around 430 cm$^{-1}$ and 800 cm$^{-1}$ \cite{Galeener2,Phillips4,McMillan,Stolen,Galeener5}. 
In our calculation within the BP approximation, the double peak found 
experimentally in the high frequency region (1100-1200 cm$^{-1}$) are better reproduced 
than in previous works \cite{Murray,Zotov}. 
Concerning the VH spectrum, Figure \protect\ref{fig:vvhBP}(b), 
three major bands can be distinguished. The double peak  
in the high frequency  region (1100-1200 cm$^{-1}$) are well placed and the curve 
is, generally, close to the experiment. The lower band ($<$ 500 cm$^{-1}$) and the intermediate
one (around 793 cm$^{-1}$) are positioned correctly but their intensities are too weak in 
comparison with the experimental lines.\\

\begin{figure}
\centerline{\includegraphics{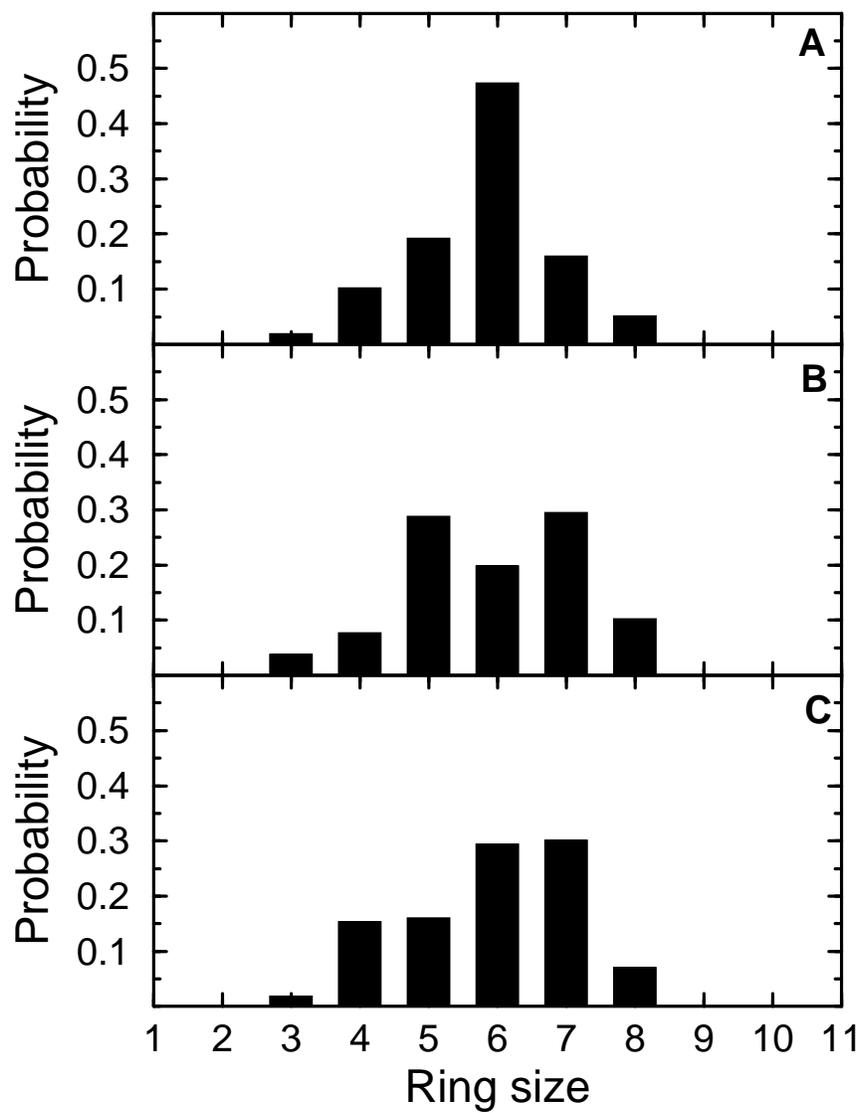}}
\caption{Distribution of the rings size in the three amorphous silica samples A, B and C.}
\label{fig:rings}
\end{figure}

In the VV Raman spectrum, the peaks around 430 cm$^{-1}$, 800 cm$^{-1}$
and 1100-1200 cm$^{-1}$ have been understood in terms of the vibrations of a CRN model
\cite{Galeener1,Barrio,Martin}. However the nature of the "defects" responsible for the D$_1$
and D$_2$ bands in the VV spectrum of a-SiO$_2$ is still the subject of a
theoretical controversy \cite{Galeener3,Phillips1,Phillips2}. As we have recalled above,
according to Galeener \cite{Galeener1,Galeener2,Galeener3,Barrio,Martin}, the D$_1$ peak
is due to the vibrational modes associated with oxygen atoms belonging to planar
fourfold rings and the D$_2$ peak to the ones associated with oxygens belonging to
threefold rings within the CRN model \cite{Zachariasen,Bell1}. In contrast to this, Phillips
\cite{Phillips1,Phillips2,Phillips3,Phillips4} argued that these bands may be associated
with skeletal surface SiO$_2$ vibrational modes in the framework of paracrystalline
theory. In the present work, the structural characteristics of the studied systems
showed, on one hand, that they contain threefold and fourfold rings and on an another
hand that, as the models of Murray and Ching \cite{Murray}, none of them contain NBO sites or free
surfaces. In Figure \protect\ref{fig:rings}, we present the different ring size distributions of
the three amorphous samples  A, B and C.
As it can be seen, the shapes of the ring size distributions are quite different for the
three samples. Since no particular trend can be derived with respect to the quench rates
used to generate these model structures,  the differences found in the distributions are
attributed to the statistics. However it is interesting to note that there is a
significant amount of three and fourfold rings in all three samples. In previous molecular-dynamics 
simulations \cite{Vollmayr}, it has been shown that the concentration of small rings 
increases with an increasing quench rate. As a consequence, the concentration of small rings in our
samples is certainly much larger than in real SiO$_2$ samples. 
We find that the number of threefold rings varies 
from 1 in samples A and C, to 2 in sample B, and the number of fourfold
rings are equal  to 4 in sample A, to 3 in sample B and to 6 in sample C.
Nevertheless, as shown in Figure \protect\ref{fig:vvhBP}, the VV Raman
spectra of our models present both D$_1$ and D$_2$ peaks which
means that the later may be correlated with the presence
of the three and fourfold rings as argued by Galeener and excludes the Phillips point of
view. \\

\subsection{Raman signature of the rings}
\label{sec:rings}

In order to clarify whether vibrational modes associated with the three and fourfold rings can be 
responsible for the D$_1$ and D$_2$ lines, we analyzed the Raman intensity due to these specific
geometries. However the connection between a given geometry in the amorphous structure and the Raman intensity
produced by it, is  not straightforward at all.

In this perspective, we carried out an analysis on the three different samples individually.
We first analyzed the modes that gave the strongest Raman activity in the 470-515 cm$^{-1}$ range
for the D$_1$ line and in the 550-730 cm$^{-1}$ range for the D$_2$ line.  The selection of these modes
was done first by  ordering the modes as a function of their Raman activity in the chosen frequency region. 
We then added together the activities of the most Raman-active modes  until the sum
reaches 80 $\%$ of the total sum: Only the modes included in those 80$\%$ were subsequently studied. 

Then, for the selected modes, 
we computed the localization entropy defined as:
\begin{equation}
S_j = - \sum_{n \delta} |< n \delta | j >|^2 \times \frac{\ln \left( |< n \delta | j >|^2 \right)}{\ln (3N)}
\end{equation}
where $|j>$ are the eigenmode components of mode $j$ and $| n \delta>$ are the $\delta$ eigenmode
components of atom $n$. The above defined entropy is connected with the lack of information
about the position of the $j$th phonon and gives the same kind of information than the 
participation ratio.  If the mode is perfectly localized, then the lack of
information is minimum and only one of the $|< n \delta | j >|^2$ is equal to one, which leads
to $S_j =0$. On the other hand, if the $j$th mode is completely delocalized, 
then $|< n \delta | j >|^2= 1/3N$, $ \forall \ n \delta$ and $S_j$ is maximum, i.e. equal to 1. 
Here $S_j$ has been normalized to 1 for convenience, thus it is not a genuine entropy.
 Let us consider the case where the $j$th mode is localized on a given ring composed of $N_R$ atoms. 
The value of $S_j$ would then be around: 
\begin{equation}
S_j^R \approx  - \sum_{n \in {\rm Ring}} \frac{1}{3N_R} \ln (\frac{1}{3N_R}) / \ln (3 N) =
\frac{\ln (3N_R)}{ \ln (3N) } 
\end{equation}
Thus if a mode is localized on all atoms in a threefold or a fourfold ring, 
the corresponding $S_j^R$ would be close to 0.529 or 0.582, respectively.
Table \protect\ref{tab:entropy} gives the localization entropies of the most 
Raman-active modes for every sample in the chosen frequency ranges. Note that $S_j$ gives
an estimation of the localization of a mode but does not tell whether a mode is localized on a {\it specific} ring.

From table \protect\ref{tab:entropy}, it is clear that the most active Raman modes in the
470-515 cm$^{-1}$ frequency range do not correspond to very localized modes since their localization
entropies lie around 0.8 and are never smaller than 0.67. The modes responsible of the D$_1$ line can therefore not
 be considered as very localized modes. For D$_2$, the situation is similar since the
localization entropies lie also between 0.72 and 0.85 for the most intensive modes in the 
550-730 cm$^{-1}$ frequency range. Therefore this criterium does not appear to be useful in order to
understand the nature  of the modes responsible for the D$_1$ and D$_2$ lines. \\

We then tried to find other criteria that could help to determine whether these modes 
can be associated to the three and fourfold rings or not. 

The first chosen criterium is based on vibrational considerations:
For every mode and for each ring, we computed the following quantity
\begin{equation}
A_{\rm Ring}(j) =  \sum_{n \in {\rm Ring}} |<n \delta | j>|^2.
\end{equation}
 If the atomic displacements due to the considered mode are of similar 
amplitude for every atom, then the
value of $A_{\rm Ring}(j)$ would be equal to 
$3 \times 8$ / 3$N$ for the fourfold rings and to $3 \times 6$ / 3$N$
for the threefold rings since there are 8 or 6  atoms included, respectively.
 We therefore computed the quantity $P_{\rm Ring}(j) = A_{\rm Ring} (j) \times 3N / 3 N_R $ where 
$N_R$=8 for the fourfold rings and $N_R$=6 for the threefold rings,  in order
to obtain an estimation of the participation of  atoms belonging to a given ring, for a given mode.
If the $j$th mode is completely localized on a ring, one obtains the maximum value for
 $P_{\rm ring}(j)$:  $P^{max}_{\rm Ring}(j) = 9.75$ for the fourfold rings 
and $P^{max}_{\rm Ring}(j) = 13$ for the 
threefold rings. Let us recall however that for a completely delocalized mode, $P_{\rm Ring}(j)$ is
equal to one and that $P_{\rm Ring}(j)=0$ means that the considered ring is not
concerned by the  $j$th mode.

The second chosen criterium is based on the contribution of a given mode and a given ring to
the total  Raman spectrum. This contribution is evaluated by computing what we name the 
"Raman ratio" which is defined in order to estimate  the contribution of each ring 
to the Raman intensity of a given mode. Let us consider a given ring. The Raman intensity 
can be divided in three contributions by considering separately the atoms belonging to the ring
and the other ones, in the following way:
\begin{eqnarray}
a_{\alpha\gamma}(j) & = &  \sum_{n  \notin {\rm Ring}}
\frac{\pi^n_{\alpha\gamma,\delta}}{\sqrt{M_n}}<n \delta
 \mid j> + \sum_{n \in {\rm Ring}}\frac{\pi^n_{\alpha\gamma,\delta}}
{\sqrt{M_n}}<n \delta  \mid j> \\  
& = & a'_{\alpha\gamma}(j) + a''_{\alpha\gamma}(j)
\end{eqnarray}

For a given polarization (we omit the Greek indices), the contribution to the 
Raman scattering can be written as:
\begin{eqnarray}
H_{\rm total}(\omega) & = & \sum_{j} a^\star_{\alpha\gamma}(j)a_{\beta\lambda}(j)\frac{1}{2 \omega_j}(\delta(\omega-\omega_j)-\delta(\omega+\omega_j)) \\
                  & = &  \sum_{j} \left[ a'_{\alpha\gamma}(j) + a''_{\alpha\gamma}(j) \right]^\star \left[a'_{\beta\lambda}(j)
+ a''_{\beta\lambda}(j) \right] \frac{1}{2 \omega_j}(\delta(\omega-\omega_j)-\delta(\omega+\omega_j)) \\
                  & = & \sum_{j} \left[ a'^\star_{\alpha\gamma}(j)a'_{\beta\lambda}(j) + a''^\star_{\alpha\gamma}(j)a''_{\beta\lambda}(j) + \left\{ a'^\star_{\alpha\gamma}(j)a''_{\beta\lambda}(j) + a''^\star_{\alpha\gamma}(j) a'_{\beta\lambda}(j)
    \right\} \right] \\ 
                   & & \times \frac{1}{2 \omega_j}(\delta(\omega-\omega_j)-\delta(\omega+\omega_j))  \\
                  & = & H_{\rm not \ Ring}(\omega) + H_{\rm Ring}(\omega) + H_{\rm overlap} (\omega)
\end{eqnarray}
To obtain the contribution  of the atoms belonging to a given ring to a selected mode $j_0$,
 we computed the integrated intensity of the spectrum,
 averaged over all possible sample orientations, in the chosen frequency range:
\begin{equation}
V^{j_0}_{\rm Ring} = V_{\rm total} - V^{j_0}_{\rm not \ Ring} 
\label{eq:Vring}
\end{equation}
with 
\begin{eqnarray}
V_{\rm total} & = & \int < H_{\rm total} (\omega) > d \omega 
\end{eqnarray}
and
\begin{eqnarray}
V^{j_0}_{\rm not \ Ring} & = & \int < H_{\rm not \ Ring}^{j_0} (\omega) > d \omega
\end{eqnarray}
where the $<...>$ denote the  average over all possible orientations and 
$H_{\rm not \ Ring}^{j_0} (\omega)$ has been obtained by setting to zero 
the $<n \delta | j_0>$ atomic components of the considered ring for the
selected mode $j_0$. The integrals run over the frequency ranges corresponding to
the D$_1$ line or to the D$_2$ line, depending on which type of rings we want
to consider (fourfold or threefold).

We then needed a quantity that could be used in order to compare the contributions
of different rings to the same mode. For this, we defined the relative Raman participation
ratio as follows. The integrated Raman intensity of the spectrum, averaged over
all possible orientations, has been computed without the contribution of mode $j_0$:
\begin{equation}
V^{j_0}_{\rm mode} = V_{\rm total} - V_{\rm not \ j_0} 
\end{equation}
where
\begin{eqnarray}
V_{\rm not \ j_0} & = & \int < H_{\rm not \ j_0} (\omega) > d \omega 
\end{eqnarray}
and
\begin{eqnarray}
H_{\rm not \ j_0} (\omega) & = &  \sum_{j \ne j_0} a^\star_{\alpha\gamma}(j)
a_{\beta\lambda}(j)\frac{1}{2 \omega_j}(\delta(\omega-\omega_j)-\delta(\omega+\omega_j)).
\end{eqnarray}
The relative Raman participation ratio is then obtained by computing the quantity:
\begin{equation}
{\rm Raman \ ratio(j_0) } = \frac{V_{\rm Ring}^{j_0}}{V_{\rm mode}^{j_0}}.
\end{equation}
The Raman ratio  gives the relative participation of the atoms
belonging to a given ring to the Raman intensity produced by the selected mode $j_0$. 

If the atoms of the chosen ring do not participate 
to the mode $j_0$ then $V_{\rm Ring}^{j_0}$ 
will be equal to zero and the Raman ratio as well. On the other hand, if only the atoms
of the ring participate to the Raman scattering, then $V_{\rm Ring}^{j_0} = V_{\rm total}$ and 
the Raman ratio is maximum, i.e. equal to $V_{\rm total}/ (V_{\rm total} - V_{\rm not \ j_0}$). 
However, from the definition of $V_{\rm Ring}^{j_0}$  (Eq. \protect\ref{eq:Vring}),
one can see that the Raman ratio can be negative: This is due to 
the fact that the $H_{\rm overlap}$ term can be negative with a magnitude
larger than that of $H_{\rm Ring}$ for the chosen ring. A negative $V_{\rm Ring}^{j_0}$
 would mean that there is a strong {\it negative} "coupling" between the atoms of the ring and the other atoms. 
A similar approach has already been used by
Lazzeri {\it et al.} for the study of the ring contributions 
to the Raman spectra of  crystalline silica phases
and zeolites and has produced negative overlaps in these phases \cite{mauri_raman}.
On the other hand, a large and positive Raman ratio would mean that the atoms of the ring participate
to the mode but also that the coupling with the other atoms might be large and {\it positive}. \\

\noindent
\subsubsection{D$_1$ line} 
\noindent
The results for the most Raman-active modes in the 470-515 cm$^{-1}$ frequency range 
are presented in table \protect\ref{tab:vibd1} 
for the A, B and C samples with the vibrational amplitudes ($P_{\rm Ring}$) on the left columns and the
Raman ratio on the right columns.

If $P_{\rm Ring}$ is larger than one for a given ring, the atoms belonging to this ring 
have larger displacements than the other atoms for the considered mode. 
From  table \protect\ref{tab:vibd1}, one can clearly see that the selected mode
always gives rise to larger displacements on atoms belonging to one (or more than one) of the fourfold rings.
As a consequence, one is tempted to  "associate" Raman-active modes to specific rings: For instance,
the 474 cm$^{-1}$ mode in sample B generates large displacements on atoms belonging to ring 1 
in that sample and the 507 cm$^{-1}$ mode in sample A generates large displacements on atoms 
belonging to rings 1 and 2 in that sample. 

The fact that some modes can be associated to several rings lead us to check whether these rings
had common atoms. The result of this analysis is presented in table \protect\ref{tab:ring}. 
Most of the time, when modes give large displacements on several rings, the later have indeed
common atoms, but this is not always true. For instance, the 500 cm$^{-1}$ mode in sample A
generates large displacements on rings 2 and 3, which do not have common atoms. This result indicates
that there exist ring-like modes which concern more than one ring.

The Raman ratio defined above has been computed for 
the most Raman-active modes in samples A, B and C and
for every fourfold ring in these systems. Results are 
presented on the right sides of table \protect\ref{tab:vibd1}
for samples A, B and C separately. We first notice that 
the Raman ratio varies approximately from  1 to -1 
and thus can be strongly negative. If we consider 
the largest positive Raman ratio, we note that they almost always
correspond  to the largest values of $P_{\rm Ring}$. 
By cross checking the largest terms in
both the vibrational amplitudes ($P_{\rm Ring}$) and the Raman ratio, 
we can easily deduce that if the Raman ratio
is large and positive, the atoms of the ring undergo 
larger displacements than the other atoms and
therefore that the contribution of these atoms to the mode is important.
On the other hand, if we consider the negative Raman ratio, 
we note that they almost always correspond
to small - or close to 1 - $P_{\rm Ring}$ values. 
This result is an indication that if the Raman ratio
is negative, then the atoms of the ring do not participate much to the mode.

\begin{figure}[ht]
\centerline{\includegraphics{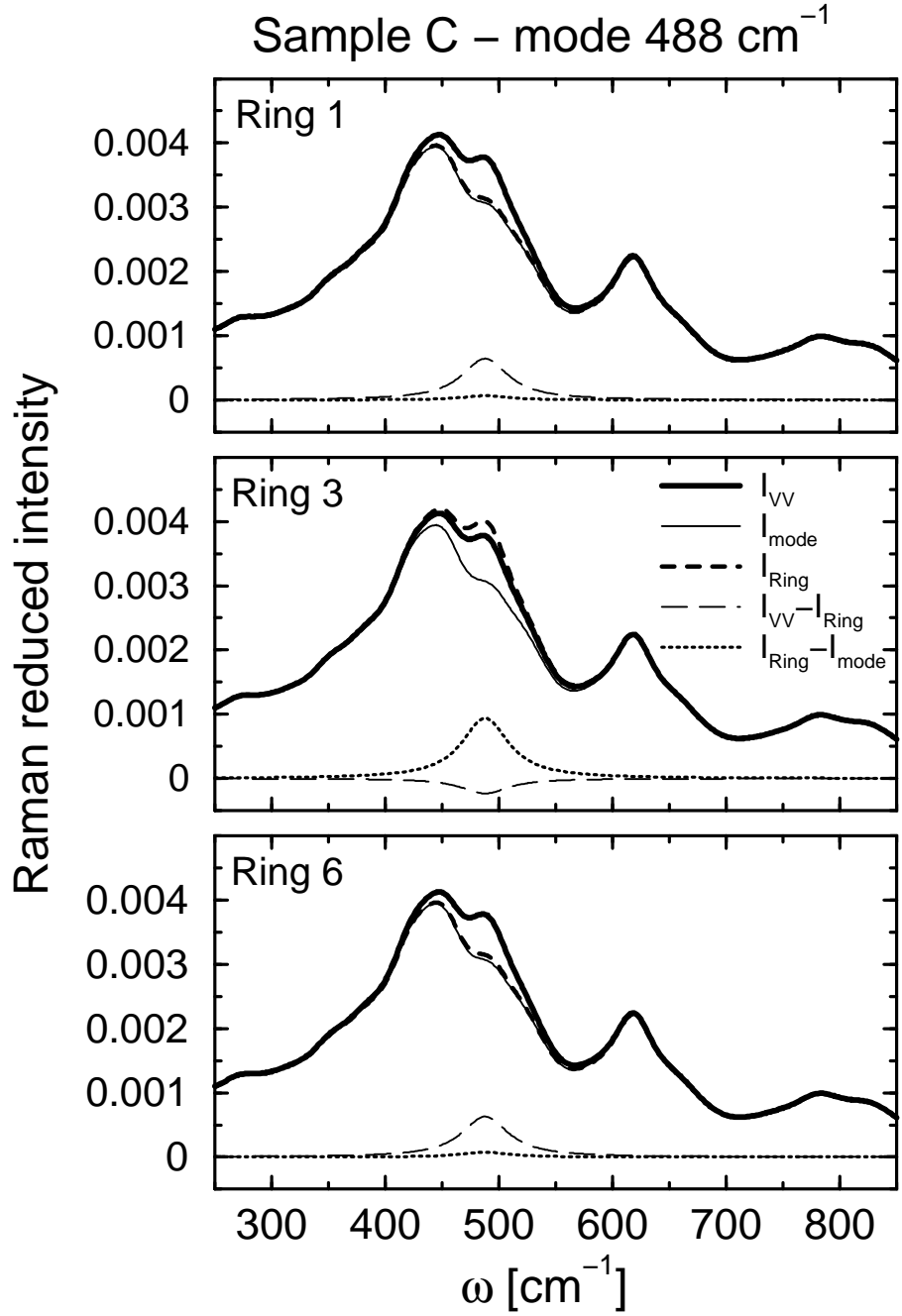}}
\caption{{\bf I$_{VV}$}: Total VV Raman spectra of sample C (bold line). {\bf I$_{\rm mode}$}:
 VV Raman spectra in which the 488 cm$^{-1}$ mode of sample C has been removed (thin line). 
{\bf I$_{\rm Ring}$}: VV Raman spectra in which the eigenmode components of atoms belonging to ring 1 (upper graph),
to ring 3 (center graph) or to ring 6 (lower graph) for the 488 cm$^{-1}$ mode of sample C
have been set to zero (dashed lines).}
\label{fig:vring_d1}
\end{figure}

In order to illustrate the results presented in tables \protect\ref{tab:vibd1}, we show
examples of Raman spectra from which the contributions of the atoms 
in individual rings have been extracted. 
Let us consider the mode at 488 cm$^{-1}$ of sample C. In figure \protect\ref{fig:vring_d1},
the total spectrum (I$_{VV}$) and the spectrum where this mode has been removed (I$_{\rm mode}$)
are depicted in bold lines. For this mode, the spectra for which the eigenmode components 
of atoms in rings 1, 3 and 6 
have been set to zero, are shown in dashed lines (I$_{\rm Ring}$). As it can be easily seen, 
for rings 1 and 6, the I$_{\rm mode}$ and I$_{\rm Ring}$ almost exactly coincide 
(I$_{\rm mode}$ - I$_{\rm Ring}$ is almost zero) and the subtractions I$_{VV}$-I$_{\rm mode}$ 
and I$_{VV}$-I$_{\rm Ring}$ are identical. This situation corresponds to a Raman ratio close to one
in table \protect\ref{tab:vibd1} and indicates that the atoms of rings 1 and 6 contribute
significantly to the 488 cm$^{-1}$ mode. On the opposite, for ring 3, the I$_{\rm Ring}$ gives
rise to a larger D$_1$ peak than that of the total VV spectrum, leading to a negative 
I$_{VV}$-I$_{\rm Ring}$ contribution and very different I$_{\rm mode}$ and I$_{\rm Ring}$ spectra. 
This indicates that the atoms of ring 3 do not participate to the 488 cm$^{-1}$ mode.

As a summary, our results suggest that the modes which produce more 
than 80 $\%$ of the Raman activity in the
frequency region of the D$_1$ line can almost always be associated to the atoms belonging to one or
more fourfold rings (i.e., the Raman ratio is larger than 0.7). Furthermore more than 70 $\%$ of the
atomic displacements due to these modes ($P_{\rm Ring}$) are located on atoms belonging to the fourfold rings. 
One exception can be found for the  \linebreak 496 cm$^{-1}$ mode in sample C, for which the Raman ratio is not larger 
than 0.5 for all rings. \\

\noindent
\subsubsection{D$_2$ line} 
\noindent
Since the D$_2$ line has been suggested to be due to modes involving the threefold rings, we analyzed
the Raman modes in the corresponding frequency region with respect to atoms belonging to such rings.
In the three SiO$_2$ samples, the threefold rings are less numerous than the fourfold ones 
(1 in sample A, 2 in sample B and 1 in sample C), hence reducing the statistics in that case.

The results for the most Raman-active modes in the 550-730 cm$^{-1}$ frequency range (corresponding to
the D$_2$ line in our spectrum) are presented in table \protect\ref{tab:vibd2} 
for samples A, B and C  with the vibrational amplitudes ($P_{\rm ring}$) on the left columns and the
Raman ratio on the right columns.
As for the D$_1$ line, we can observe that there is a correlation between the largest $P_{\rm Ring}$ values
and the largest positive Raman ratio. We can also note that not all the Raman-active
modes in the D$_2$ line region correspond to modes associated to the threefold rings, whereas the
contrary is true: All rings participate to, at least, one Raman-active mode in the D$_2$ line region.

\begin{figure}[ht]
\centerline{\includegraphics{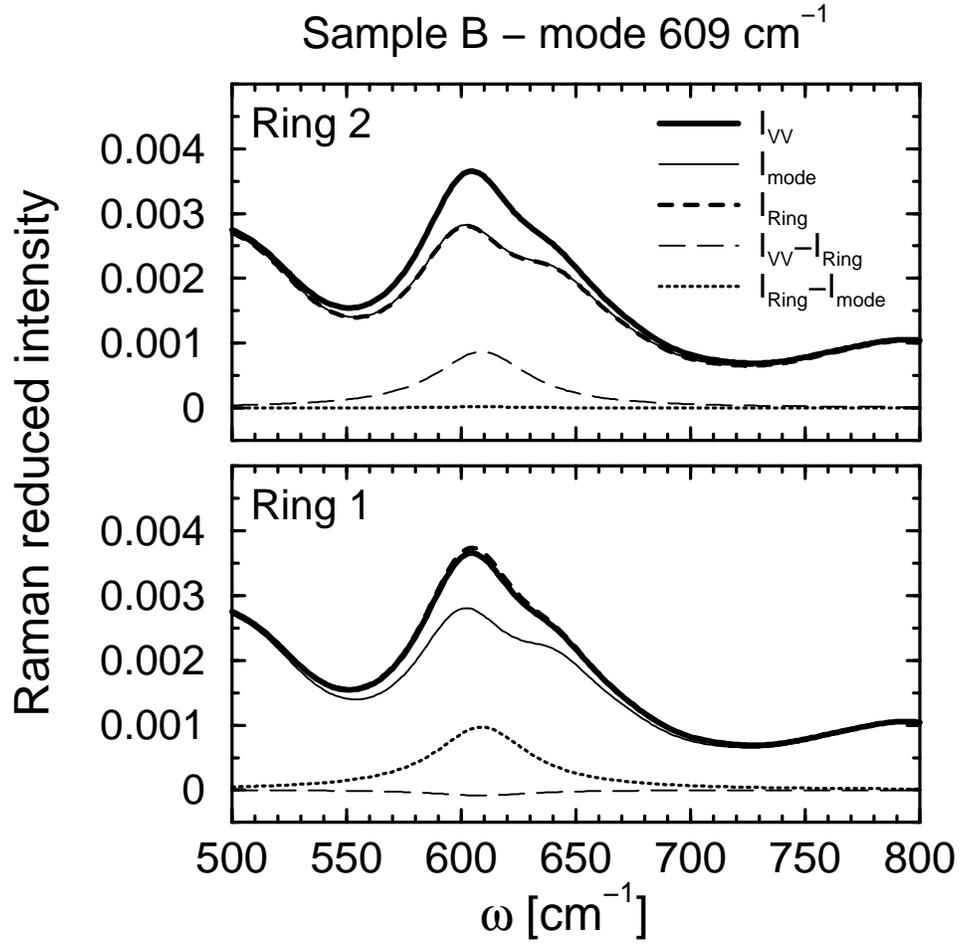}}
\caption{{\bf I$_{VV}$}: Total VV Raman spectra of sample B (bold line). {\bf I$_{\rm mode}$}:
 VV Raman spectra in which the 609 cm$^{-1}$ mode of sample B has been removed (thin line). 
{\bf I$_{\rm Ring}$}: VV Raman spectra in which the eigenmode component of atoms belonging to ring 2 (upper graph),
or to ring 1 (lower graph) for the 609 cm$^{-1}$ mode of sample B
have been set to zero (dashed lines). }
\label{fig:vring_d2}
\end{figure}

As for the D$_1$ line, we illustrate the results presented in table \protect\ref{tab:vibd2} by the
Raman spectra depicted in figure \protect\ref{fig:vring_d2}. In this figure, we chose to show
the Raman spectrum in which the mode of frequency 609 cm$^{-1}$ in sample B has been removed 
(I$_{\rm mode}$)
together with the Raman spectra in which the eigenmode components 
of atoms belonging to the threefold rings have been set to zero (I$_{\rm Ring}$ for rings 1 and 2).
It is clear that if the eigenmode components of atoms in ring 2 are set to zero, the Raman spectrum
is identical to the one obtained if the selected mode is removed,
 i.e. the resulting difference I$_{\rm Ring}$
- I$_{\rm mode}$ (dotted line) is zero.
 On the opposite, the contribution of atoms of ring 1 
to the 609 cm$^{-1}$ mode is non-existent.

Recently Umari and Pasquarello \cite{Umari2} have deduced the proportion of
oxygen atoms belonging to the threefold and fourfold rings in an experimental sample
 using a direct relation between the areas under the experimental and calculated 
 D$_1$ and D$_2$ peaks, respectively.
In view of the results presented in tables \protect\ref{tab:vibd1} and \protect\ref{tab:vibd2}, 
such an approach seems a bit hazardous. Indeed, we have demonstrated that among the most
Raman-active modes in the regions of the D$_1$ and D$_2$ lines (i.e. modes that produce
more than 80 $\%$ of the peaks), some can not be associated to modes involving 
atoms of a fourfold or a threefold ring. This result means that 
there is no proportionality relation between the area under the peaks and 
the proportion of threefold and  fourfold rings in the SiO$_2$ glass.

\section{Raman scattering in compressed amorphous SiO$_2$}

\subsection{Generation of the compressed samples}

In order to mimic the effect of pressure, the box size of the amorphous sample C has
been rescaled to three different sizes in successive steps. First, the box length and
coordinates were rescaled from the experimental density of 2.20 g/cm$^3$ to a density of
2.40 g/cm$^3$ (sample C1) and a Car-Parrinello molecular-dynamics simulations  of 1.9 ps
was launched at 300 K, using the CPMD code \cite{CPMD}. 
 The entire procedure was repeated for two other
densities: 2.67 g/cm$^3$ (sample C2) and 4.28 g/cm$^3$ (sample C3). For the sample at
2.67 g/cm$^3$, the initial coordinates were taken
 and rescaled from the final ones of
the  simulation performed on sample C1, and for the sample at 4.28 g/cm$^3$,
the initial coordinates were taken and rescaled from the final ones of the 
simulation performed on sample C2.  In this latter case, an important modification of the
structure leading to an large increase of the ionic temperature was observed. Therefore during
the first ps of the {\it ab initio} MD simulation, the ionic temperature has been constrained to
stay at 300 K. This constraint was then released for the rest of the {\it ab initio} MD
simulation on sample C3. 
In order to achieve good convergence for the calculation of the pressures in samples
C1, C2 and C3, higher energy cutoffs would be required.
We estimated that this information was not worth the computational effort 
and obtained the pressures from the volume to pressure conversion using the experimental
 equation of state of a-SiO$_2$ \cite{Hemley}. The pressures of our samples are thus evaluated to
$\approx$ 3 GPa for sample C1, $\approx$ 7 GPa for sample C2 and $\approx$ 34 GPa for sample C3.

\begin{figure}
\centerline{\includegraphics{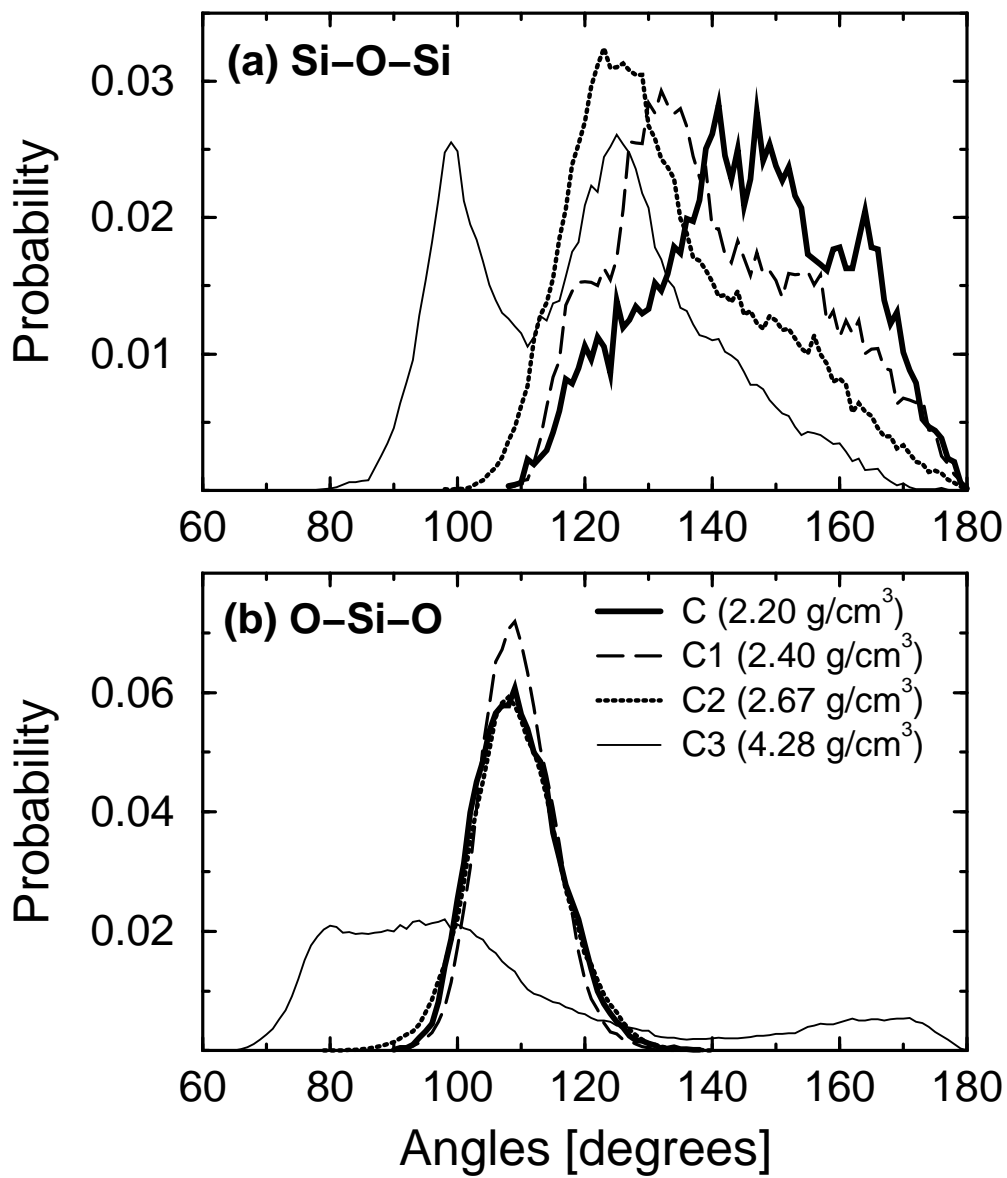}}
\caption{Distribution of the Si-O-Si angles (a) and of the O-Si-O angles (b) 
of sample C at different densities: C at 2.20 g/cm$^3$ (bold line), 
C1 at 2.45 g/cm$^3$ (dashed line), C2 at 2.67 g/cm$^3$ (dotted line) and 
C3 at 4.28 g/cm$^3$ (solid line). }
\label{fig:angles_comp}
\end{figure}

For  samples C1 and C2, we observe no modifications of the network topology. 
The observed structural modifications only concern the shift of
the Si-O-Si angle distributions towards smaller angles (Figure
\protect\ref{fig:angles_comp}) and no changes in the Si-O bond lengths and
in the ring statistics are noticed. In contrast to this, the structure of sample C3
is very different from the ones of sample C, C1 and C2:
The tetrahedral network is destroyed and replaced by a mixing of four, five  and sixfold-coordinated 
silicon atoms connected by corners and/or edges, and threefold-coordinated oxygen atoms. These coordination
changes explain the very different Si-O-Si and O-Si-O angular distributions observed in
Figure \protect\ref{fig:angles_comp}.

\begin{figure}
\centerline{\includegraphics{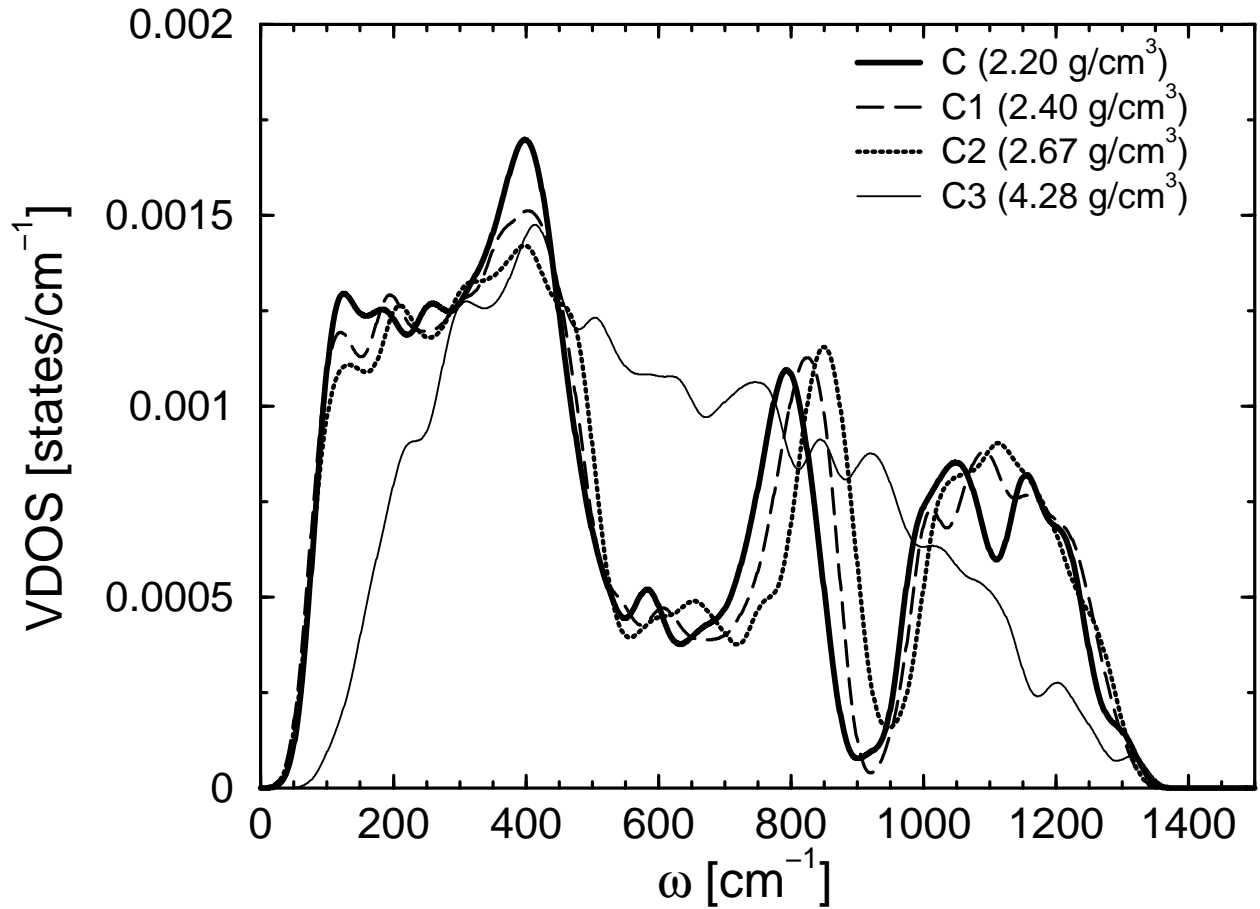}}
\caption{Vibrational density of states of the sample C at different densities: 
C at 2.20 g/cm$^3$ (bold line), C1 at 2.45 g/cm$^3$ (dashed line), C2 at 2.67 g.cm$^3$ 
(dashed-dotted line) and C3 at 4.28 g/cm$^3$ (solid line). A Gaussian broadening of 
$2\sigma$ = 35 cm$^{-1}$ has been used.}
\label{fig:DOS_comp}
\end{figure}

At the end of these {\it ab initio} molecular-dynamics simulations, the atomic
coordinates were relaxed to 0 K and the dynamical matrices were
computed as described in section \protect\ref{sec:sio2}.
 For every sample, 
we then obtained the vibrational frequencies and the corresponding modes by a diagonalisation of
the dynamical matrices.
The vibrational densities of states of sample C and of the three compressed samples C1, C2 and C3
are depicted in Figure \protect\ref{fig:DOS_comp}.
As pressure is increased from C to C2, the double peak in the high frequency region disappears
and the peaks at  $\sim$600 cm$^{-1}$ and $\sim$800 cm$^{-1}$ are shifted to higher wavenumbers. 
One can also notice a decrease of the peak intensity at $\sim$400 cm$^{-1}$. 
On the other hand, the vibrational density of states of sample C3 exhibits very different
features which reflect the changes in the structure: No clear gap can be seen in the broad
peak going from $\sim$100 cm$^{-1}$ up to 1300 cm$^{-1}$.

\subsection{Raman spectra and signature of the rings} 
\begin{figure}
\centerline{\includegraphics{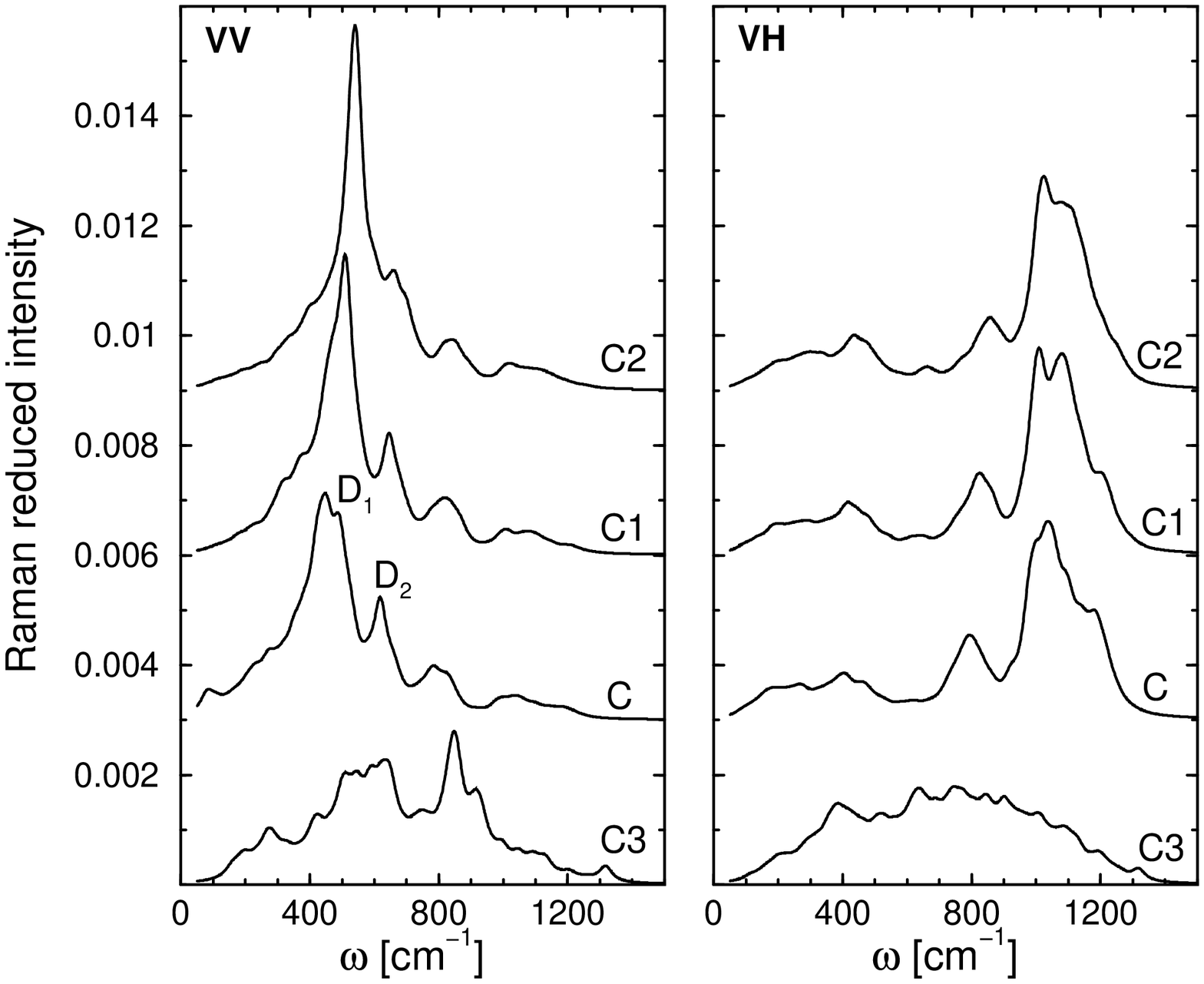}}
\caption{Frequency dependence of the normalized VV Raman reduced intensity (left panel) and of the VH 
Raman intensity (right panel) for the sample
C at different densities:  C at 2.20 g/cm$^3$, 
C1 at 2.45 g/cm$^3$, C2 at 2.67 g/cm$^3$ and 
C3 at 4.28 g/cm$^3$. The spectra have been shifted vertically for clarity.}
\label{fig:Raman_comp}
\end{figure}

\noindent
In Figure \protect\ref{fig:Raman_comp}, the VV (left panel) and VH (right panel) 
Raman spectra of the compressed samples C1, C2 and C3 are compared to that of sample C. 
Firstly, we note that the VV and VH spectra of  sample C3 are very different 
compared to that of samples C, C1 and C2, which is due to its
very different density of states and structure. For the slightly compressed samples (C1 and C2)
the main band and the D$_1$ line in  the VV spectra  merge into a sharp main peak.
 The intensity of this peak increases with increasing density and moves to higher wavenumbers.
The increasing sharpness of the band around D$_1$ with increasing density could be attributed to the 
shift to lower angle and the increasing sharpness of the Si-O-Si angle distribution 
shown in Figure \protect\ref{fig:angles_comp}. 
In contrast with this behavior, the D$_2$ line intensity does not increase with
increasing density, however its position slightly shifts towards higher wavenumbers. 
One can also note that all the bands lower than 900 cm$^{-1}$ clearly shift to higher wavenumbers. 
On the opposite, in the high frequency region, 
we observe that the double peak slightly shifts to lower wavenumbers.
Overall these behaviors have been observed 
experimentally \cite{McMillan,Hemley} and reproduced by other numerical studies 
\cite{Murray}. For the VH spectra, we observe that the principal bands shift to higher 
wavenumbers except for the largest peak of the double peak in the high frequency region. \\

In order to show the effect of compression on the ring contributions, we have evaluated the 
localization entropies, the vibrational amplitudes $P_{\rm Ring}$ and the Raman ratio as defined
in section \protect\ref{sec:rings} for all rings in samples C1 and C2. The sample C3 was
not studied in that respect since the tetrahedral network is damaged in that sample.  
The results are presented in table \protect\ref{tab:entropycomp} for the entropies and in
table \protect\ref{tab:vibcompd1} and table \protect\ref{tab:vibcompd2} 
for the $P_{\rm Ring}$ values and the Raman ratio in the frequency regions of the
D$_1$ and D$_2$ lines, respectively.

The results presented in table  \protect\ref{tab:entropycomp} show that the modes 
contributing to the sharp line around 500 cm$^{-1}$ are less numerous in the compressed samples C1 and C2
than in sample C. For samples C1 and C2, the frequency range in which these modes 
have been sought for was modified to take into account the frequency shift of the 
peaks. These frequency regions have been set to 490-540 cm$^{-1}$ for sample C1 and
to 505-585 cm$^{-1}$ for sample C2. It should be noted that the intense sharp peak at 540 cm$^{-1}$ 
in sample C2 is due to  only two modes that produce 80 $\%$ of the total Raman intensity.
However, the entropies of these modes are not much different from the ones of the modes 
responsible of the D$_1$ line in the uncompressed samples.
A similar conclusion on the localization entropy can be given 
for what concerns the most Raman-active modes in
the frequency region around the D$_2$ line (605-770 cm$^{-1}$). Note however that there
are more Raman-active modes in this region than in the D$_2$ region of the uncompressed
sample C.

The vibrational amplitudes ($P_{\rm Ring}$) and the Raman ratio 
(see definitions in section \protect\ref{sec:rings})
were then evaluated for the selected modes in the frequency regions of interest. 
The results for the modes in the 490-540 cm$^{-1}$ and 505-585 cm$^{-1}$ frequency regions
 are presented in table \protect\ref{tab:vibcompd1}. 
The mode at 503 cm$^{-1}$ in sample C1 is clearly the same mode
than the 488 cm$^{-1}$ one in sample C since the $P_{\rm Ring}$ values  and the Raman ratio
are maximum on  rings 1 and 6 for this mode. The mode at 512 cm$^{-1}$ in sample C1 
could also be related to the 496 cm$^{-1}$ mode in sample C if one considers the
value of $P_{\rm Ring}$  which is maximum on ring 5. Upon compression these two modes have been shifted of 
roughly 15 cm$^{-1}$. However, all the other modes that were associated to one or two rings in sample C, 
have disappeared in sample C1 and, upon further compression in sample C2, even the modes that were 
still present in sample C1 vanished. For sample C2, the remaining modes in the  505-585 cm$^{-1}$ frequency 
range do not present large values of $P_{\rm Ring}$ or of the Raman ratio: 
This indicates that the Raman-active modes in this frequency region can not be associated
to motions of atoms belonging  to a fourfold ring.
Since the same rings of sample C still exist in sample C2, this result means that the compression
induced a structural modification of the rings and of 
their environment that produced the extinction of the ring-like modes in the Raman spectra. 
When sample C is compressed into C1 and C2, the network connections (and thus the rings) do not change 
with increasing density but only the angles and relative distances between atoms do. 
The changes in the Raman spectra are therefore not due to a
different proportion of  rings but only to changes in the structure of the rings, 
mainly the decrease of the Si-O-Si angles. 
A more detailed analysis showed that the ring-like modes still exist in this
frequency range (i.e. there are modes for which
$P_{\rm Ring}$ is maximum for a given fourfold ring) but
their Raman activities are much smaller, if not zero.

The effect of compression seems to be a little different for the modes in the frequency
region corresponding to the D$_2$ line. The $P_{\rm Ring}$ values and the Raman ratio
for the 605-770 cm$^{-1}$ frequency range are presented in table \protect\ref{tab:vibcompd2}
for the compressed samples C1 and C2. In both samples, one can see that there
is always one mode being clearly associated to the threefold ring: The 646 cm$^{-1}$ mode
in sample C1 and the 664 cm$^{-1}$ one  in sample C2. This result indicates that the mode at
619 cm$^{-1}$ in sample C is still present in samples C1 and C2 but is shifted of approximately 
28 cm$^{-1}$ and 45 cm$^{-1}$, respectively, upon compression.
The ring-like mode associated to the threefold ring remains therefore Raman-active upon
larger compression than the ones associated to the fourfold rings. This is certainly due
to the fact that the Si-O-Si angles in the threefold rings have less possibility to
change than in the fourfold rings. Indeed we checked that the Si-O-Si angles of the
threefold ring in sample C, C1 and C2 change by less than 3$^{\circ}$ upon compression,
whereas changes of more than 20$^{\circ}$ were noticed in the case of the Si-O-Si angles
of the fourfold rings. \\

\section{Conclusion}
We have presented results for VV and VH Raman spectra for uncompressed and compressed
amorphous silica models generated by \emph{ab initio} molecular-dynamics simulations. 
To model scattering in these materials 
we have considered the bond-polarizability mechanism. This mechanism gives rise to VV 
and VH spectra in good agreement with experimental data 
in shapes and peak positions. 
Analysis of the VV partial Raman spectra confirms that the D$_1$ and D$_2$ peaks
 are due to the presence of four and threefold rings respectively, as suggested
by Galeener. The association of ring-like modes to the peaks was done by using
two approaches, one based on the atomic motions of the atoms belonging to
the ring and the other one based on the calculation of the partial Raman spectra
for atoms belonging to the ring. Both approaches give qualitatively the same results and show
that not all  rings necessarily produce a strong Raman activity.

 The Raman spectra obtained for compressed samples were also computed and are in
very good agreement with experimental results by Hemley  {\it et al.} \cite{Hemley} for densified samples. 
As long as the tetrahedral network is maintained, 
we find that the Raman peak around 500 cm$^{-1}$ is not necessarily due to 
modes associated to the fourfold rings although these rings are still present
in the glass sample. 
However the D$_2$ line still exists upon compression and can still be  associated
to the threefold rings. 

These results suggest that (i) there exist collective ring-like modes (involving several rings),
(ii) the presence of the D$_1$ and D$_2$
lines can be associated to the presence of fourfold and threefold rings
in the glass sample, (iii) the absence of the D$_1$ and D$_2$
lines does not mean that there are no fourfold and threefold rings in the sample, 
(iv) the area under the D$_1$ and the D$_2$ peaks are not simply proportional 
to the concentration of small rings in the SiO$_2$ sample, as it has been previously proposed.

As a conclusion, we were able to show that a combination of classical models 
(the molecular-dynamics simulations for the generation of the amorphous samples and 
the scattering models) and of first-principles calculations (the {\it ab initio} 
computation of the dynamical matrices) can be used 
to successfully compute the Raman spectra of amorphous silica and to extract 
detailed information about the Raman signature of structural units such as small rings. \\

{\bf Acknowledgements}\\

We thank  Walter Kob for very helpful discussions 
and  a careful reading of the manuscript.
The computations were performed at CINES (Montpellier, France) on  IBM SP3 computer. The
work was supported by a CNRS-France/CNCPRST-Morocco agreement.%

\newpage

\begin{table}
\caption{Frequency, Raman intensity and localisation entropy for the most Raman-active modes in the 470-515 cm$^{-1}$
 frequency range (D$_1$ line) and in the 550-730 cm$^{-1}$ frequency range (D$_2$ line) 
for the three SiO$_2$ samples A, B and C.}

\centerline{D$_1$ line}
\begin{tabular}{|c|c|c||c|c|c||c|c|c|}
\hline
\multicolumn{3}{|c||}{sample A} & \multicolumn{3}{c||}{sample B} & \multicolumn{3}{c|}{sample C} \\  \hline
frequency & Raman & localisation & frequency & Raman & localisation & frequency & Raman & localisation  \\
(cm$^{-1}$) & intensity  &  entropy  & (cm$^{-1}$) & intensity &  entropy & (cm$^{-1}$) & intensity  &  entropy \\ 
 & (a.u.) & & & (a.u.) & & & (a.u.) & \\ \hline 
507 & 5.76$\cdot$10$^{-3}$ & 0.801 & 498 & 5.32$\cdot$10$^{-3}$ &  0.806 & 406 & 7.16$\cdot$10$^{-3}$ & 0.831 \\
500 & 4.86$\cdot$10$^{-3}$ & 0.804 & 474 & 4.38$\cdot$10$^{-3}$ & 0.678 & 488 & 6.76$\cdot$10$^{-3}$ & 0.777 \\
474 & 4.42$\cdot$10$^{-3}$ & 0.757 & 512 & 4.37$\cdot$10$^{-3}$ & 0.811 & 486 & 5.36$\cdot$10$^{-3}$ & 0.804 \\
       &                &          & 489 & 2.09$\cdot$10$^{-3}$ & 0.836 & 520 & 3.96$\cdot$10$^{-3}$ & 0.823 \\
       &                &          & 521 & 1.87$\cdot$10$^{-3}$ & 0.847 & 501 & 2.20$\cdot$10$^{-3}$ & 0.872 \\
       &                &          & 486 & 1.77$\cdot$10$^{-3}$ & 0.847 &         &       &  \\

\hline
\end{tabular}
\vspace{0.5cm}

\centerline{D$_2$ line}
\begin{tabular}{|c|c|c||c|c|c||c|c|c|}
\hline
\multicolumn{3}{|c||}{sample A} & \multicolumn{3}{c||}{sample B} & \multicolumn{3}{c|}{sample C} \\  \hline
frequency & Raman & localisation & frequency & Raman & localisation & frequency & Raman & localisation  \\
(cm$^{-1}$) & intensity &  entropy  & (cm$^{-1}$) & intensity &  entropy & (cm$^{-1}$) & intensity  &  entropy \\ 
 & (a.u.) & & & (a.u.) & & & (a.u.) & \\ \hline 
634 &  7.61$\cdot$10$^{-3}$ &  0.81   &  602   &  1.05$\cdot$10$^{-2}$ & 0.82 & 619 & 1.53$\cdot$10$^{-2}$ & 0.75 \\
661 &  2.88$\cdot$10$^{-3}$ &  0.82   &  609   &  7.28$\cdot$10$^{-3}$ & 0.73 & 652 & 2.82$\cdot$10$^{-3}$ & 0.85  \\
606 &  1.68$\cdot$10$^{-3}$ &  0.82   &  639   &  6.91$\cdot$10$^{-3}$ & 0.72 & 665 & 2.12$\cdot$10$^{-3}$ & 0.79 \\
639 &  1.64$\cdot$10$^{-3}$ &  0.85   &  596   &  5.51$\cdot$10$^{-3}$ & 0.83 &        &                & \\   
\hline
\end{tabular}

\label{tab:entropy}
\end{table}

\begin{table}
\caption{$P_{\rm Ring}$ and Raman ratio (see text for definitions) for the most Raman-active modes
in the 470-515 cm$^{-1}$ frequency range for the three SiO$_2$ samples A, B and C.} 

\centerline{Sample A}
\begin{tabular}{|c||c|c|c|c||c|c|c|c|}
\hline
Frequency  & \multicolumn{4}{c||}{$P_{\rm Ring}$} & \multicolumn{4}{c|}{Raman ratio} \\
 (cm$^{-1}$) & ring 1 & ring 2 & ring 3 & ring 4 & ring 1 & ring 2 & ring 3 & ring 4 \\ \hline 
507 & 3.16 & 2.35 & 0.76 & 1.35  & 0.83 & 0.70 & -0.32 & 0.42 \\
500 &  1.11 & 2.60 & 2.63 & 0.40 & 0.14 & 0.59 & 0.76 & 0.21 \\
474 &  0.30 & 0.41 & 0.44 & 1.96 & 0.08 & 0.15 & 0.16 & 0.33 \\
\hline
\end{tabular}

 \vspace*{0.5cm}
\centerline{Sample B} 
\begin{tabular}{|c||c|c|c||c|c|c|}
\hline
Frequency & \multicolumn{3}{c||}{$P_{\rm Ring}$} & \multicolumn{3}{c|}{Raman ratio} \\ 
 (cm$^{-1}$) & ring 1 & ring 2 & ring 3 & ring 1 & ring 2 & ring 3 \\ \hline 
498 &  0.39 & 0.63 & 3.81 & 0.04 & 0.23 & 0.89  \\
474 &  5.24 & 0.35 & 0.16 & 0.95 & -0.04 & -5 10$^{-3}$ \\
513 &  0.47 & 2.99 & 0.85 & 0.19 &  0.89 & -0.43 \\
489 &  0.61 & 1.23 & 1.31 & 0.11 & -0.52 & 0.79 \\
\hline
\end{tabular}

\vspace*{0.5cm}
\centerline{Sample C}
\begin{tabular}{|c||c|c|c|c|c|c||c|c|c|c|c|c|}
\hline
Frequency & \multicolumn{6}{c||}{$P_{\rm Ring}$} & \multicolumn{6}{c|}{Raman ratio}   \\ 
 (cm$^{-1}$) & ring 1 & ring 2 & ring 3 & ring 4 & ring 5 & ring 6 &  ring 1 & ring 2 & ring 3 & ring 4 & ring 5 & ring 6 \\ \hline
496 & 0.50 & 0.80 & 0.98 & 1.12 & 2.62 & 0.53 & 0.22 & 0.09 & -0.08 & 0.18 & 0.49 & 0.23 \\
488 & 3.99 & 0.18 & 0.76 & 0.67 & 0.53 & 3.93 & 0.91 & 0.18 & -0.33 & -0.29 & 0.05 & 0.89 \\
486 & 0.79 & 1.59 & 2.32 & 2.55 & 1.55 & 0.61 & 0.46 & 0.66 & 0.68 & 0.73 & -0.52 & 0.43 \\
520 & 0.67 & 1.37 & 2.88 & 1.17 & 2.48 & 0.32 & 0.30 & -0.83 & 0.99 & 0.56 & 0.90 & 0.23 \\
\hline
\end{tabular}

\label{tab:vibd1}
\end{table}

\begin{table}
\caption{Number of common atoms in the fourfold rings of samples A, B,  C, C1 and C2.}
\begin{tabular}{|c||cccc||ccc||cccccc|}
\hline
 & \multicolumn{4}{c||}{Sample A} & \multicolumn{3}{c||}{Sample B} & \multicolumn{6}{c|}{Samples C, C1 and C2} \\
 &  ring 1 & ring 2 & ring 3 & ring 4 & ring 1 & ring 2 & ring 3 & ring 1 & ring 2 & ring 3 & ring 4  & ring 5 & ring 6 \\
\hline
ring 1 &  - & 3 & 0 & 0 &  - & 1 & 0 &        - & 0 & 0 & 0 & 0 & 3 \\
ring 2 &  3 & - & 0 & 0 & 1 & - & 0 &        0 & - & 0 & 0 & 0 & 0  \\
ring 3 &  0 & 0 & - & 0 &  0 & 0 & - &        0 & 0 & - & 3 & 3 & 0 \\
ring 4 &  0 & 0 & 0 & - &    &   &   &         0 & 0 & 3 & - & 1 & 0 \\
ring 5 &    &   &   &   &    &   &   &         0 & 0 & 3 & 1 & - & 0 \\
ring 6 &    &   &   &   &    &   &   &         3 & 0 & 0 & 0 & 0 & - \\
\hline

\end{tabular}

\label{tab:ring}
\end{table}

\begin{table}
\caption{$P_{\rm Ring}$ and Raman ratio (see text for definitions) for the most Raman-active modes
in the 550-730 cm$^{-1}$ frequency range  for the three SiO$_2$ samples A, B and C.}

\centerline{Sample A} 
\begin{tabular}{|c||c||c|}
\hline
Frequency  & $P_{\rm Ring}$ \ \ \ & Raman ratio \ \ \ \\
 (cm$^{-1}$) & ring 1 & ring 1  \\ \hline 
634 & 3.33 & 0.83 \\
661 & 3.16 & 0.97 \\
606 & 0.69 & 0.24 \\
639 & 1.71 & 0.83  \\
\hline
\end{tabular}

 \vspace*{0.5cm}
\centerline{Sample B} 
\begin{tabular}{|c||c|c||c|c|}
\hline
Frequency & \multicolumn{2}{c||}{$P_{\rm Ring}$ \ \ } & \multicolumn{2}{c|}{Raman ratio \ \ } \\ 
 (cm$^{-1}$) & ring 1 & ring 2 &  ring 1 & ring 2\\ \hline 
602 & 1.47 & 1.74 & 0.29 & 0.55 \\
609 & 0.98 & 6.24 & -0.09 & 0.98  \\
639 & 5.52 & 1.43 & 0.99 & -0.62  \\
596 & 0.62 & 1.08 & 0.42 & 0.38 \\
\hline
\end{tabular}

\vspace*{0.5cm}
\centerline{Sample C}

\begin{tabular}{|c||c||c|}
\hline
Frequency & $P_{\rm Ring}$ \ \ \ & Raman ratio \ \ \ \\ 
 (cm$^{-1}$) & ring 1 &  ring 1  \\ \hline
619 & 5.65 & 0.88 \\
652 & 1.49 & 0.78  \\
665 & 0.58 & 0.54 \\
\hline 
\end{tabular}

\label{tab:vibd2}
\end{table}

\begin{table}
\caption{{\bf Upper table:} 
Frequency, Raman intensity and localisation entropy of the most Raman-active
modes in the 490-540 cm$^{-1}$ frequency range for sample C1 and in the 505-585 cm$^{-1}$ frequency
range for sample C2. 
{\bf Lower table:} Frequency, Raman intensity and localisation entropy of the most Raman-active
modes in the 605-770 cm$^{-1}$ frequency range for samples C1 and C2.}

\centerline{D$_1$ line}
\begin{tabular}{|c|c|c||c|c|c|}
\hline
\multicolumn{3}{|c||}{Sample C1} & \multicolumn{3}{c|}{Sample C2} \\
\hline 
Frequency & Raman & Localisation & Frequency & Raman & Localisation  \\
(cm$^{-1}$) & intensity (a.u.) & entropy & (cm$^{-1}$) & intensity (a.u.) & entropy\\ \hline
512 & 3.57$\cdot$10$^{-3}$ & 0.86 & 546 & 3.42$\cdot$10$^{-2}$ & 0.85 \\
503 & 6.52$\cdot$10$^{-3}$ & 0.81 & 534 & 2.15$\cdot$10$^{-2}$ & 0.86 \\
\hline
\end{tabular}

\vspace*{0.5cm}
\centerline{D$_2$ line}
\begin{tabular}{|c|c|c||c|c|c|}
\hline
\multicolumn{3}{|c||}{Sample C1} & \multicolumn{3}{c|}{Sample C2} \\
\hline 
Frequency & Raman & Localisation & Frequency & Raman & Localisation \\
(cm$^{-1}$) & intensity (a.u.) & entropy & (cm$^{-1}$) & intensity (a.u.) & entropy \\ \hline
646 & 1.54$\cdot$10$^{-2}$ & 0.79 & 664 & 1.48$\cdot$10$^{-2}$ & 0.79 \\
677 & 4.56$\cdot$10$^{-3}$ & 0.83 & 698 & 9.65$\cdot$10$^{-3}$ & 0.83 \\
632 & 2.42$\cdot$10$^{-3}$ & 0.83 & 643 & 5.89$\cdot$10$^{-3}$ & 0.87 \\
       &                &  & 608 & 5.80$\cdot$10$^{-3}$ &  0.80 \\
\hline
\end{tabular}

\label{tab:entropycomp}
\end{table}
 
\begin{table}
\caption{$P_{\rm Ring}$ and Raman ratio (see text for definitions) for the most
Raman-active modes in the 490-540 cm$^{-1}$ frequency range for sample C1 (upper table)
and in the 505-585 cm$^{-1}$ frequency range for sample C2 (lower table). }

\centerline{Sample C1}
\begin{tabular}{|c||c|c|c|c|c|c||c|c|c|c|c|c|}
\hline
Frequency & \multicolumn{6}{c||}{$P_{\rm Ring}$} & \multicolumn{6}{c|}{Raman ratio}   \\ 
 (cm$^{-1}$) & ring 1 & ring 2 & ring 3 & ring 4 & ring 5 & ring 6 &  ring 1 & ring 2 & ring 3 & ring 4 & ring 5 & ring 6 \\ \hline
512 & 1.15 & 1.15 & 1.43 & 1.28 & 2.22 & 0.58 & 0.30 & 0.26 & 0.30 & 0.18 & 0.27 & 0.21 \\
503 & 3.35 & 0.32 & 1.02 & 0.86 & 0.78 & 3.17 & 0.82 & 0.22 & -0.39 & -0.32 & 0.16 & 0.81 \\
\hline
\end{tabular} 

\vspace*{0.5cm}
\centerline{Sample C2}
\begin{tabular}{|c||c|c|c|c|c|c||c|c|c|c|c|c|}
\hline
Frequency & \multicolumn{6}{c||}{$P_{\rm Ring}$} & \multicolumn{6}{c|}{Raman ratio}   \\ 
 (cm$^{-1}$) & ring 1 & ring 2 & ring 3 & ring 4 & ring 5 & ring 6 &  ring 1 & ring 2 & ring 3 & ring 4 & ring 5 & ring 6 \\ \hline
546 & 0.78 & 0.50 & 0.88 & 1.22 & 0.73 & 1.71 & 0.20 & 0.17 & 0.19 & 0.24 & 0.03 & 0.34 \\
534 & 1.85 & 0.24 & 0.85 & 1.39 & 1.09 & 1.00 & 0.40 & -3 10$^{-3}$ & 0.22 & 0.13 & 0.23 & 0.24 \\
\hline
\end{tabular}

\label{tab:vibcompd1}
\end{table}

\begin{table}
\caption{$P_{\rm Ring}$ and Raman ratio (see text for definitions) for the most
Raman-active modes in the 605-770 frequency range for samples C1 (upper table)
and C2 (lower table).}

\centerline{Sample C1}

\begin{tabular}{|c||c||c|}
\hline
Frequency & $P_{\rm Ring}$ & Raman ratio  \\ 
 (cm$^{-1}$) & ring 1 &  ring 1  \\ \hline
646 & 4.83 & 0.85 \\
677 & 1.13 & 0.48   \\
632 & 1.18 & 0.72 \\
\hline
\end{tabular}

\vspace*{0.5cm}
\centerline{Sample C2}
\begin{tabular}{|c||c||c|}
\hline
Frequency & $P_{\rm Ring}$ & Raman ratio  \\ 
 (cm$^{-1}$) & ring 1 &  ring 1  \\ \hline
664 & 3.53  & 0.79 \\
698 & 0.65  & -0.21  \\
643 & 0.98  & 0.41 \\
608  & 1.13 & 0.51 \\
\hline
\end{tabular}

\label{tab:vibcompd2}
\end{table}

\newpage

\begin{thebibliography}{999}


\bibitem{Galeener1} F.L. Galeener, Phys. Rev. B {\bf 19}, 4292 (1979).
\bibitem{Galeener2} F.L. Galeener, Solid State Commun. {\bf 44}, 1037 (1982).
\bibitem{Galeener3} F.L. Galeener, R.A Barrio, E. Martinez, and R.J. Elliott, Phys. Rev.
Lett. {\bf 53}, 2429 (1984).
\bibitem{Barrio} R.A Barrio, F.L. Galeener, and E. Martinez, Phys. Rev. Lett. {\bf 52},
1786 (1984).
\bibitem{Martin} R. M. Martin and F.L. Galeener, Phys. Rev. B {\bf 23}, 3071 (1981).
\bibitem{Zachariasen} W. H. Zachariasen , J. Am. Chem. Soc. {\bf 54}, 3841(1932).
\bibitem{Bell1} R. J. Bell and P. J. Dean, Phil. Mag. {\bf 25}, 1381(1972).
\bibitem{Phillips1} J. C. Phillips, in Solid State Physics, edited by Ehrenreich, F.
Seitz, and D. Turnbull (Academic, New York, 1982), vol. {\bf 37}, p.93.
\bibitem{Phillips2} J. C. Phillips, J. Non-Cryst. Solids {\bf 63}, 347 (1984).
\bibitem{Phillips3} J. C. Phillips, Phys. Rev. B {\bf 32}, 5350 (1985).
\bibitem{Phillips4} J. C. Phillips, Phys. Rev. B {\bf 33}, 4443 (1986).
\bibitem{Bell2} R. J. Bell, Methods in computational Physics (Academic, New York,
1976), {\bf 15}, p.215.
\bibitem{Murray} R. A. Murray, and W. Y. Ching, Phys. Rev. B {\bf 39}, 1320 (1989).
\bibitem{Zotov} N. Zotov, I. Ebbsj\"o, D. Timpel, and H. Keppler, Phys. Rev. B {\bf 60},
6383 (1987).
\bibitem{Umari} P. Umari, A. Pasquarello, and A. Dal Corso, Phys. Rev. B {\bf 63}, 94305
(2001).
\bibitem{Umari2}
P. Umari and A. Pasquarello, Physica B {\bf 316-317}, 572 (2002).
\bibitem{vanBeest} B. W. H. van Beest,  G. J. Kramer, and R. A. van Santen, Phys. Rev.
Lett. {\bf 64}, 1955 (1990).
\bibitem{Benoit1} M. Benoit and W. Kob, Europhys. Lett. {\bf 60}, 269 (2002).
\bibitem{rahmani} A. Rahmani, J.-L. Sauvajol, S. Rols and C. Benoit, Phys. Rev. B {\bf 66}, 125404 (2002).
\bibitem{viliani} G. Viliani {\it et al.}, Phys. Rev. B {\bf
52}, 3346 (1995);
Phys.: Condens. Matter {\bf 9}, 2149 (1997).
\bibitem{CPMD} CPMD V3.4 Copyright IBM Corp. 1990-2001, 
Copyright MPI f\"ur Festk\"orperforschung Stuttgart 1997-2001.
\bibitem{Taraskin}
S.N. Taraskin and S.R. Elliott, Phys. Rev. B {\bf 55}, 117 (1997).
\bibitem{Carpenter} J.M. Carpenter and D.L. Price, Phys. Rev. Lett. {\bf 54}, 441 (1985).
\bibitem{wischnewski98} A. Wischnewski, U. Buchenau, A. J. Dianoux, W. A. Kamitakahara, and
J. L. Zarestky, Phys. Rev. B {\bf 57}, 2663 (1998).

\bibitem{McMillan} P. McMillan, B. Piriou, and R. Couty, J. Chem. Phys. {\bf 81}, 4234
(1984).
\bibitem{Stolen} R. H. Stolen and G. E. Walrafen, J. Chem. Phys. {\bf 64}, 2623 (1976).
\bibitem{Galeener5} A.E. Geissberger and F.L. Galeener, Phys. Rev. B {\bf 28}, 3266 (1983).
\bibitem{Vollmayr}
K. Vollmayr, W. Kob and K. Binder, Phys. Rev. B {\bf 54}, 15808 (1996).
\bibitem{mauri_raman} M. Lazzeri and F. Mauri, Phys. Rev. Lett. {\bf 90}, 036401 (2003).
\bibitem{Hemley} R. J. Hemley, H. K. Mao, P. M. Bell, and B. O. Mysen, Phys. Rev.
Lett. {\bf 57}, 747 (1986).

\end{thebibliography}
\end{document}